\documentclass[aps,prl,twocolumn,superscriptaddress,]{revtex4-2}
\usepackage{cancel}
\usepackage[normalem]{ulem}
\usepackage{bm}
\usepackage{amsmath}
\usepackage{amsthm}
\usepackage{natbib}
\usepackage{amssymb}
\usepackage{graphicx}

\makeatletter
\usepackage{times}
\usepackage{comment}
\usepackage{graphicx}
\usepackage{feynmf}
\usepackage{tabularx}
\usepackage{amsmath}
\usepackage{amstext}
\usepackage{amssymb}
\usepackage[hidelinks,colorlinks=true,urlcolor=blue,linkcolor=blue,citecolor=blue]{hyperref}
\usepackage{amsfonts}
\usepackage{longtable,booktabs}
\usepackage{url}
\usepackage{subfigure}
\usepackage{amsbsy}
\usepackage{amsthm}
\usepackage{bm}
\usepackage{cleveref}
\usepackage{mathrsfs}
\usepackage{amsfonts}
\usepackage{amsbsy}
\usepackage{color}

\def\bphi{{\boldsymbol \phi}}
\def\brho{{\boldsymbol \brho}}
\def\bk{{\boldsymbol k}}
\def\bd{{\boldsymbol d}}

\def\bp{{\boldsymbol p}}

\def\br{\boldsymbol{r}}

\def\bj{{\boldsymbol j}}
\def\bs{{\boldsymbol s}}
\def\bx{{\boldsymbol x}}

\def\bu{{\boldsymbol u}}
\def\bv{{\boldsymbol v}}
\def\bPhi{{\boldsymbol \Phi}}

\def\bphi{{\boldsymbol \phi}}

\def\bL{{\boldsymbol L}}

\def\bM{{\boldsymbol M}}
\def\bV{{\boldsymbol V}}
\def\bfu{\bm {\mathfrak {u}}}
\def\la{\langle}

\def\calPT{\mathcal{PT}}

\def\calM{\mathcal{M}}
\def\calN{\mathcal{N}}

\def\calD{\mathcal{D}}

\def\calE{\mathcal{E}}
\def\calJ{\mathcal{J}}

\def\e{\epsilon}

\def\pa{\partial}
\def\nn{\nonumber}

\def\la{\langle}
\def\ra{\rangle}
\begin{document}
\title{Unconventional spin Hall effect in $\mathcal {PT}$ symmetric spin-orbit coupled quantum gases}
\author{Hui Tang}
\altaffiliation{These authors contributed equally to this work.}
\affiliation{Shenzhen Institute for Quantum Science and Engineering, Southern University of Science and Technology, Shenzhen 518055, China}
\author{Guan-Hua Huang}
\altaffiliation{These authors contributed equally to this work.}
\affiliation{Hefei National Laboratory, Hefei 230088, China}
\author{Shizhong Zhang}
\affiliation{Department of Physics and HKU-UCAS Joint Institute for Theoretical and Computational Physics at Hong Kong, University of Hong Kong, Hong Kong, China}
\author{Zhongbo Yan}
\email{yanzhb5@mail.sysu.edu.cn}
\affiliation{Guangdong Provincial Key Laboratory of Magnetoelectric Physics and Devices, School of Physics, Sun Yat-sen University, Guangzhou 510275, China}
\author{Zhigang Wu}
\email{wuzhigang@quantumsc.cn}
\affiliation{Quantum Science Center of Guangdong-Hong Kong-Macao Greater Bay Area (Guangdong), Shenzhen 508045, China}

\date{\today }
\begin{abstract}
We theoretically study the intrinsic spin Hall effect in $\mathcal {PT}$ symmetric, spin-orbit coupled quantum gases confined in an optical lattice. The interplay of the $\mathcal {PT}$ symmetry and the spin-orbit coupling leads to a doubly degenerate non-interacting band structure in which the spin polarization and the Berry curvature of any Bloch state are opposite to those of its degenerate partner. Using experimentally available systems as examples, we show that such a system with a two-component Fermi gas exhibits an intrinsic spin Hall effect akin to that found in the context of electronic materials. For a two-component Bose gas, however, an unconventional spin Hall effect emerges in which the spin polarization and the currents are coplanar and the spin Hall conductivity displays a characteristic anisotropy.  We propose to detect such an unconventional spin Hall effect in harmonically trapped systems using dipole oscillations and perform extensive numerical simulations to validate the proposal. Our work paves the way for quantum simulation of the solid-state intrinsic spin Hall effect and experimental explorations of unconventional spin Hall effects in quantum gases. 
\end{abstract}
\maketitle
\emph {Introduction.---}Quantum simulations using ultracold atomic gases are often intended to shed light on other complex systems that are less accessible experimentally~\cite{Bloch2012, Gross2017,TARRUELL2018,Bohrdt2021}. They have also inadvertently led to the discovery of novel quantum phenomena, owing to the versatility of atomic systems.~\cite{Bloch2008,Cooper2019}. For example, consider the simulation of the electronic spin-orbit coupling (SOC), which plays a crucial role in various solid-state phenomena, including anomalous Hall effects, spin Hall effects, and topological insulators and superconductors~\cite{Manchon2015,Bihlmayer2022}.  The creation of synthetic SOC for ultracold atoms~\cite{Lin2011,Wang2012} is no doubt motivated by the desire to better understand these phenomena and the spin-orbit coupled Fermi gas has indeed become a fertile ground for studying many of them~\cite{ZhangJ2014}. However, with the additional choice of bosonic atoms and the ability to engineer synthetic SOC beyond the traditional solid-state types~\cite{Galitski2013,Zhai2015}, it has also led to the realization of fascinating phases of matter beyond the condensed matter paradigms, such as supersolids~\cite{Li_ss_2017} and various topologically non-trivial superfluids~\cite{Zhang_soc_2018}. 

Among many SOC-driven phenomena, the spin Hall effect (SHE), i.e., the generation of a transverse spin current by an electric field, occupies an important place due to its potential applications in spintronics and electronic devices~\cite{DYAKONOV1971, Hirsch1999,Sinova2004,Murakami2003,Sinova2015}. Shortly after the synthetic SOC was realized~\cite{Lin2011}, the SHE was demonstrated in a $^{87}$Rb gas with an effective 1D SOC~\cite{Beeler2013}. This 1D SOC is equivalent to an Abelian gauge potential, which generates a spin-dependent Lorentz force responsible for the SHE~\cite{Zhu2006,Liu2007}. This pioneering experiment serves as the simplest conceptual example of SHE in a quantum gas but is not a simulation of the intrinsic SHE discussed in the context of 2D electronic materials. In the latter systems, the 2D SOC amounts to a non-Abelian gauge potential~\cite{Jin_2006,Hatano2007} and generates finite Berry curvatures of the Bloch bands which, along with the dynamics of the spin degree of freedom, underpins the intrinsic SHE~\cite{Culcer2004,Culcer2005}.  Since most experimental observations of SHE in electronic materials involve contributions from both the intrinsic and the impurity-related extrinsic mechanisms, an unequivocal experimental verification of the former is generally difficult in those systems~\cite{Vignale2009}. Therefore, a quantum simulation of the intrinsic SHE in a spin-orbit coupled Fermi gas free from any impurity effect would be of great interest and would further expand the capabilities of these systems in the study of spintronics~\cite{Aidelsburger2013,Kennedy2013,LiNatu2014,Oshima2016,Armaitis2017}.  More importantly, a fundamental question concerns whether an intrinsic SHE exists in a spin-orbit coupled Bose gas and, if it does, whether it displays any basic character distinct from its fermionic counterpart. 

In this Letter, we show that $\mathcal {PT}$ symmetric 2D spin-orbit coupled quantum gases, recently made available experimentally~\cite{Sun2018,Liang2023}, are ideal systems to address these questions. The non-interacting band structure of such a system is endowed with geometric properties that are conducive to the SHE. Indeed, for the Fermi gas this system realizes a close analog of models used in the theoretical studies of the intrinsic SHE of electronic materials~\cite{Murakami2006}. For the Bose gas, however,  an unconventional intrinsic SHE can emerge in the ground state phase in which the spins are polarized in the 2D plane. In contrast to the Fermi gas, such an in-plane magnetization breaks the rotation symmetry of the system and gives rise to a unique anisotropy in the spin Hall conductivity. We perform extensive numerical simulations to demonstrate that this unconventional intrinsic SHE can be observed experimentally by means of dipole oscillations~\cite{Wu2011,Wu2014,Wu2015,Anderson2019}. Our work thus lays the theoretical foundation for quantum simulation of the solid-state intrinsic SHE and experimental exploration of unconventional SHE in quantum gases. 

\emph{ SHE in $\calPT$ symmetric quantum gases.---}We first present an heuristic picture of how the SHE can arise in a two-component spin-orbit coupled quantum gas confined in a 2D optical lattice and possessing the $\calPT$ symmetry. It is well-known that the presence of  the $\calPT$ symmetry in a lattice system leads to a double degeneracy for each Bloch band characterized by the band index $n$ and the quasi-momentum $\bk$~\cite{Armitage2018}.  It is convenient to span this degenerate subspace using a basis that diagonalizes the spin $s_z=\frac{1}{2}\sigma_z $. We denote these basis states as $\bphi_{n\bk} = (\phi_{n\bk\uparrow}, \phi_{n\bk\downarrow})^{T}$ and  $\bphi_{\bar n\bk} =  (\phi_{\bar n\bk\uparrow}, \phi_{\bar n\bk\downarrow})^T$, which are related to each other by the $\calPT$ symmetry, i.e., $\bphi_{n\bk} = \calPT \bphi_{\bar n\bk} $. From this relation and the properties of the $\calPT$ symmetry we  find that $\la\bphi_{n\bk} | s_z |\bphi_{n\bk}\ra = - \la\bphi_{\bar n\bk}| s_z |\bphi_{\bar n\bk}\ra $ and $\Omega_{n}(\bk) =- \Omega_{\bar n}(\bk) $, where $\Omega_{n}(\bk) \equiv  -2{\rm Im} \la \pa_{k_x}\bfu_{n\bk}| \pa_{k_y}\bfu_{n\bk} \ra$ is the Berry curvature of the cell-periodic Bloch state $\bfu_{n\bk} = e^{-i\bk\cdot \br} \bphi_{n\bk} $. Namely the spin polarization and the Berry curvature of any Bloch state are opposite to those of its degenerate partner. Now, if an external force is applied to the system, an atom occupying the Bloch state $\bphi_{ n\bk}$ will gain a transverse anomalous velocity proportional to the Berry curvature $\Omega_{n}(\bk)$~\cite{Xiao2010}.  The association of the spin polarization and the Berry curvature then implies that the atoms with opposite spin polarizations will move in opposite transverse directions. 

For a two-component  degenerate Fermi gas in which the atoms populate the degenerate states evenly, this leads immediately to the SHE in a charge-neutral system, i.e., the generation of a transverse spin current with net zero transverse mass current. Remarkably, it can also happen to a Bose gas in which the atoms tend to condense into a single lowest energy state. This is because, as we shall see shortly,  the competition of the atomic interactions and the SOC can compel the atoms to condense in the superposition of the two degenerate band minimums with equal weight. However, unlike the degenerate Fermi gas which has no net magnetization, the Bose condensate in such a case is a coherent state with spin polarized in the 2D plane. 

\emph {Spin conductivity tensor.---}To further contrast the intrinsic SHE in the Fermi and Bose gases,  we examine the spin conductivity tensor 
\begin{align}
\sigma^{s}_{\mu\nu} = -\lim_{\omega\rightarrow 0} \frac{1}{A\omega} {\rm Im}\chi^{s}_{\mu\nu}(\omega),
\label{kubo}
\end{align}
where $ A$ is the area of the quasi-2D system, and $\chi^{s}_{\mu\nu}(\omega)$ is the Fourier transform of the retarded spin current-current correlation function 
$
    \chi^{s}_{\mu\nu}(t-t') = -i\theta(t-t')\langle[\hat J_\mu^{s}(t),\hat J_\nu(t')]\rangle
$; here $\hat J_\mu $ and  $\hat J^{s}_\mu$ are, respectively,  the total mass current and spin current operator. Under the rotation of the 2D coordinates $\br' = \hat R(\theta)\br$, i.e., $x' = x\cos\theta + y\sin\theta $ and $y' = - x\sin\theta  + y\cos\theta $, the spin conductivity tensor transforms as
$
    \sigma_{\mu'\nu'}^{s}=\sum_{\mu,\nu}R_{\mu'\mu}(\theta)R_{\nu'\nu}(\theta)\sigma_{\mu\nu}^{s}
$, or more specifically 
\begin{align}
    \sigma_{y'x'}^{s}&=\frac{1}{2}\sin2\theta(\sigma_{yy}^{s}-\sigma_{xx}^{s})+\cos^2\theta\sigma_{yx}^{s}-\sin^2\theta\sigma_{xy}^{s},  \nn \\
\sigma_{x'x'}^{s}&=\frac{1}{2}\sin 2\theta(\sigma_{yx}^{s}+\sigma_{xy}^{s})+\cos^2\theta\sigma_{xx}^{s}+\sin^2\theta\sigma_{yy}^{s}. 
\label{scttr}
\end{align}
We shall see that a fundamental difference between the SHE in the Fermi and Bose gases is reflected by the rotational behavior of the spin conductivity tensor.

\begin{figure}[b]
\begin{centering}
\includegraphics[width=8.6cm]{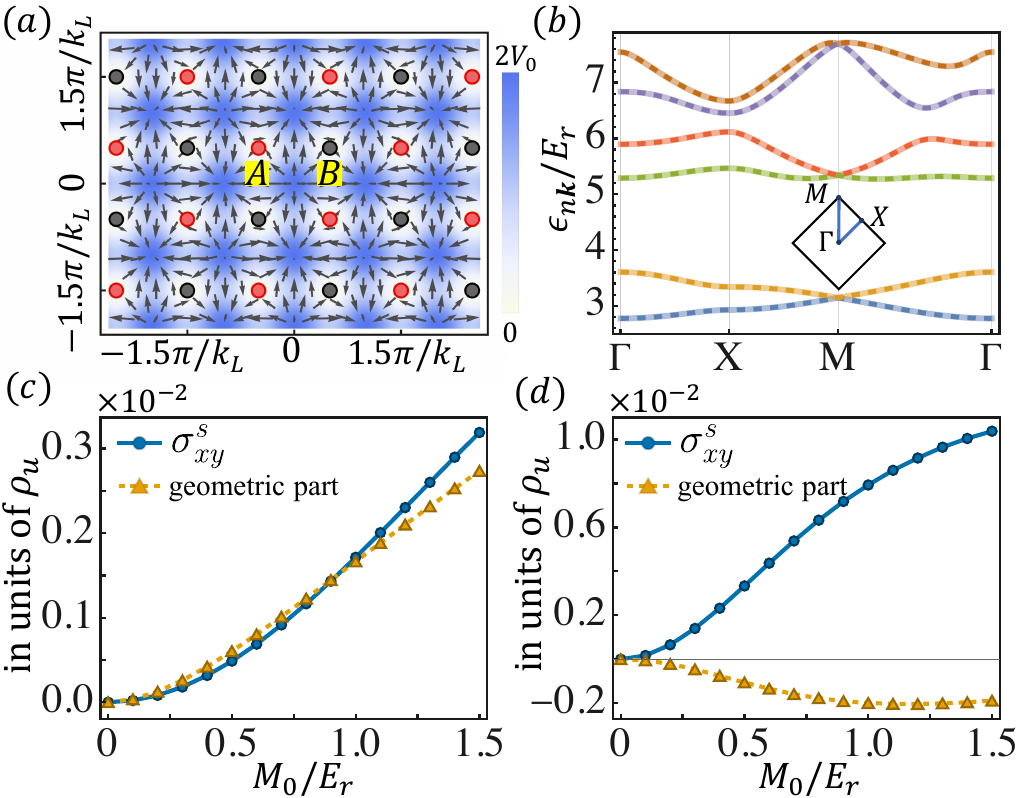}
\par\end{centering}
\caption{(a) Illustration of the optical lattice and the Raman potentials; the latter, depicted by the arrows, can be viewed as a fictitious in-plane magnetic field $\bV_{\rm R} = (V_{\rm R1}, V_{\rm R2})$ with nonzero divergences; (b) The doubly degenerate non-interacting bands; (c) and (d) The spin Hall conductivity and its geometric part for the Fermi gas in an insulating state (c) with a unit cell density $\rho_u = 4$ and in a metallic state (d) with $\rho_u = 3$. Here $V_0 = 4E_r$ and $M_0 = 1.5 E_r$, where $E_r = k_L^2/2m$ is the recoil energy ($\hbar  = 1$ throughout the paper).}
\label{fig:fermi}
\end{figure}

We proceed to calculate this quantity for experimental $\calPT$ symmetric spin-orbit coupled quantum gases, which are recently realized with $^{87}$Sr for the degenerate Fermi gas~\cite{Liang2023} and $^{87}$Rb for the Bose condensate~\cite{Sun2018}. Both are confined in a 2D square optical lattice potential $V_{\text{latt}}(\br)=V_0(\cos^2k_{L}x+\cos^2k_{L}y)$, where the 2D SOC is created by two Raman lattice potentials $ V_{\rm R1}(\br)=2M_0\sin (k_{L}x)\cos (k_{L}y)$ and $V_{\rm R2}(\br) = 2M_0\sin (k_Ly)\cos (k_{L}x)$, as illustrated in Fig.~\ref{fig:fermi}. The single-particle Hamiltonian for such a system is
 \begin{equation}
	\label{h0}
	h=\left[{\bm p^2}/{2m}  +V_{\text{latt}}(\br) \right ]  +V_{\rm R 1}(\br) s_x + V_{\rm R2}(\br)s_y,
\end{equation}
where  $s_x = \frac{1}{2}\sigma_{x}$, $s_y =\frac{1}{2} \sigma_{y}$ and the identity matrix in spin space is suppressed. Note that the square lattice of the optical potential is divided into two sublattices, $A$ and $B$, distinguished by the local Raman potentials around the lattice sites. It can be easily checked that the Hamiltonian  indeed possesses the $\calPT$ symmetry which, as mentioned previously, results in the double degeneracy of the energy bands $\e_{n\bk} $.  This is also  explicitly verified by solving $h \bphi_{n\bk} = \e_{n\bk} \bphi_{n\bk}$ (see Fig.~\ref{fig:fermi}(b)). 

We first consider the degenerate Fermi gas at zero temperature. The total mass (spin) current operator is 
$
\hat J^{(s)}_\mu = \sum_ {nn'\bk} j^{(s)}_{\mu,nn'}(\bk)  \hat a^\dag_{n\bk}\hat a_{n'\bk},
$
where $\hat a_{n'\bk}$ annihilates an atom occupying $\bphi_{n\bk}$;  $ j_{\mu,nn'}(\bk)  = \la \bphi_{n\bk}| j_\mu |\bphi_{n'\bk}\ra $ and $ j^{s}_{\mu,nn'}(\bk)  = \la \bphi_{n\bk}|j^{s}_\mu |\bphi_{n'\bk}\ra $ are the matrix elements of the single-particle mass current $ \bj = \bp/m$ and spin current  $ \bj^{s} = \frac{1}{2}\left \{ s_z, \bj \right \}$ respectively. Thus one finds from Eq.~(\ref{kubo}) 
\begin{align}
    \sigma^{s}_{\mu\nu}=\frac{1}{A}\sum_{\bk,n\neq n'}(f_{n\bk}-f_{n'\bk})\frac{{\rm Im} [j^{s}_{\mu,nn'}(\bk)j^*_{\nu,nn'}(\bk)]}{(\e_{n\bk}-\e_{n'\bk})^2},
    \label{fermisigma}
\end{align}
where $f_{n\bk} = \theta(\mu_{F}-\e_{n\bk})$ is the Fermi-Dirac distribution and $\mu_{F}$ is the chemical potential. We make two observations about Eq.~(\ref{fermisigma}). First, the ground state of the Fermi gas clearly respects the $C_4$ rotation symmetry of the Hamiltonian, which indicates that
$
\sigma_{xx}^{s}=\sigma_{yy}^{s}$ and $ \sigma_{xy}^{s}=-\sigma_{yx}^{s}$. These conditions, when substituted in Eq.~(\ref{scttr}), immediately leads to the isotropy of the spin Hall conductivity $\sigma^{s}_{xy}$. Second, if we approximate the matrix elements of the spin current operator by $j^{s}_{\mu,nn'}(\bk) \approx \la \bphi_{n\bk}| s_z |\bphi_{n\bk}\ra  j_{\mu,nn'}(\bk)$, we find that $ \sigma^{s}_{xy} \approx -\frac{1}{A}\sum_{n\bk} f_{n\bk} \la \bphi_{n\bk}| s_z |\bphi_{n\bk}\ra  \tilde\Omega_n(\bk)$, where $\tilde\Omega_n(\bk) = -2{\rm Im} \la \pa_{k_x}\bfu_{n\bk}|(1-\hat P) | \pa_{k_y}\bfu_{n\bk} \ra $ with $\hat P \equiv  \sum_{n'\neq n}f_{n'\bk} |\bfu_{n'\bk} \ra \la \bfu_{n'\bk} |$. The quantity $\tilde\Omega_n(\bk) $ is intimately related to the Berry curvature $\Omega_n(\bk) $ and also possesses the property that $\tilde\Omega_n(\bk)  =- \tilde\Omega_{\bar n}(\bk) $ for $\calPT$ symmetrical partner states. Such an approximation amounts to retaining only the geometric part of the spin Hall conductivity and is a formal justification of our earlier arguments about the Berry curvature induced SHE. However, it neglects the contribution from the dynamics of the spin degree of freedom, i.e., those from the spin dipole and spin torque~\cite{Culcer2004}. Interestingly, as shown in Fig.~\ref{fig:fermi}(c) and (d), we find that the geometric part of $ \sigma^{s}_{xy} $ is an excellent approximation for insulating states while the spin dipole and spin torque play a dominant role in spin transport for metallic states. 

\begin{figure}[tb]
\begin{centering}
\includegraphics[width=8.6cm]{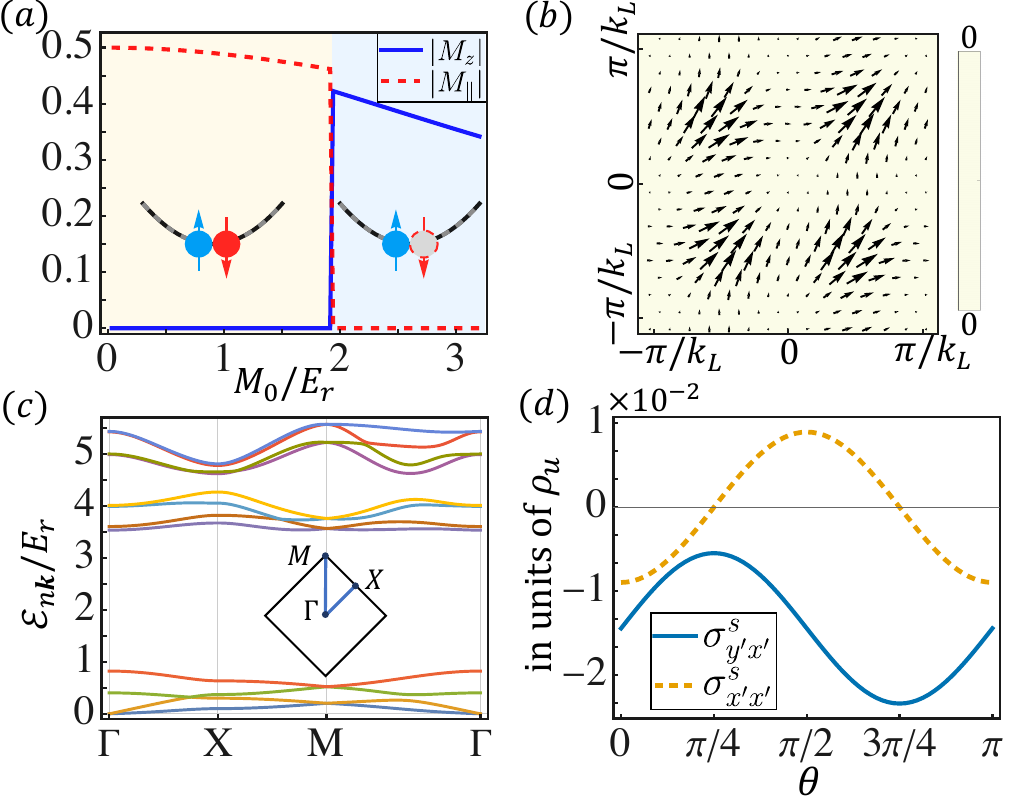}
\par\end{centering}
\caption{(a) The spin-orbit coupled Bose condensate undergoes a transition from a perpendicularly magnetized phase to an in-plane magnetized one as the SOC strength varies. Here $V_0 = 8E_r$, $\rho g_{\uparrow\uparrow} = 0.26 E_r $ and $\rho g_{\uparrow\downarrow} = 0.22 E_r$, where $\rho = N/A$ is the cell-averaged atomic density. (b) The magnetization density  of the condensate mode $\bPhi_1$ in an area containing four lattice sites, where the uniform background color indicates a vanishing $z$-component.  Here $M_0 = 1.5 E_r$. (c) The Bogoliubov spectrum corresponding to the ground state in (b). (d) The spin conductivity tensor $\sigma^{s}_{\mu\nu}$ as a function of the rotation angle $\theta$ of the coordinates. Here $\rho_u$ is the number of atoms per unit cell.}
\label{fig:bose}
\end{figure}

We examine next the two-component Bose gas, for which the atomic interactions between the spin components are essential in determining its ground state.  We consider the case of anisotropic interactions found in experiments, where the interaction strengths $g_{\sigma\sigma'}$ satisfies $g_{\uparrow\uparrow}= g_{\downarrow\downarrow} > g_{\uparrow\downarrow}$. The ground state spinor condensate wave function $\bPhi(\br) = (\Phi_{\uparrow}(\br), \Phi_\downarrow(\br))^T$, normalized to unity,  is determined by the spinor Gross-Pitaevskii (GP) equation~\cite{Kawaguchi2012}
\begin{align}
\sum_{\sigma'}[h_{\sigma\sigma'}\Phi_{\sigma'}+N g_{\sigma\sigma'}|\Phi_{\sigma'}|^2\Phi_{\sigma}]=\mu_{B}\Phi_{\sigma},
\label{GPequ}
\end{align}
where $N$ is the atom number  and $\mu_{B}$ is the chemical potential. By solving this equation for various SOC strength $M_0$,  two quantum phases can be found for this system~\cite{Chen2023}, distinguished by the nature of the magnetization $ \bM= \la \bPhi |\bs|\bPhi \ra$, where $\bs = (s_x,s_y,s_z)$. For sufficiently large $M_0$, the atoms condense at one of the two degenerate band minimums at the $\Gamma$ point, indicating a two-fold degeneracy in the condensate mode. Since the spins of these two Bloch states are oppositely polarized along the $z$-direction, the condensate forms a perpendicularly magnetized state with $\bM = \pm |M_z| \hat {\boldsymbol z}$ which breaks a $Z_2$ symmetry (see Fig.~\ref{fig:bose}(a)); an anomalous Hall effect was shown to exist in this phase~\cite{Huang2022,Chen2023}. As $M_0$ decreases, the atoms are forced into a coherent supposition of the two minimums at the $\Gamma$ point as a result of the competition between SOC and atomic interactions, leading to an in-plane magnetization. In this case, there is a four-fold degeneracy in the condensate mode because of the four possible choices in the relative phase $\theta_l$  between the two degenerate states, where $\theta_l =  {l\pi}/{2}-{\pi}/{4}$ with $l = 1,\cdots,4$~\cite{SM}. Each of these four modes, denoted by $\bPhi_l$, is differentiated by the direction of its corresponding in-plane magnetization $\bM= |M_\parallel| \hat\br_l $, where $\hat\br_l = \cos \theta_l \hat {\boldsymbol x}+\sin \theta_l \hat {\boldsymbol y}$.   The atoms spontaneously select one of these four modes to condense at and thereby breaks the $C_4$ symmetry (see Fig.~\ref{fig:bose}(b)).  Importantly, the condensate mode $\bPhi_l$ retains a subset of the full space symmetry group of the Hamiltonian. In particular, it is still invariant under the $\pi$ rotation operation $\calD_l = e^{-i\pi (\bs+\bL)\cdot \hat\br_l}$ around the $\hat \br_l$ axis,  where $\bL$ is the single particle angular momentum operator. As we shall see, this property plays a critical role in determining the angular dependence of the spin Hall conductivity. 

To compute the spin conductivity tensor for the in-plane magnetized phase we first obtain the elementary excitations of the condensate by solving the spinor Bogoliubov-de Gennes (BdG) equations~\cite{Huang2021}
\begin{align}
(h+ \calM -\mu_{B} )\bu_{n\bk} - \calN \bv_{n\bk} &= \calE_{n\bk}\bu_{n\bk};\nn \\
-(h^*+ \calM^*  -\mu_{B} )\bv_{n\bk} + \calN^* \bu_{n\bk} &= \calE_{n\bk}\bv_{n\bk},
\end{align}
where $\calE_{n\bk}$ is the energy of the quasi-particle, $\bu_{n\bk}$ and $\bv_{n\bk}$ are the corresponding spinor Bogoliubov amplitudes,
${\mathcal M}_{\sigma\sigma'} = N[g_{\sigma\sigma'}\Phi^*_\sigma \Phi_{\sigma'} +\delta_{\sigma\sigma'}\sum_{\sigma''} g_{\sigma\sigma''}|\Phi_{\sigma''}|^2]$ and $\calN_{\sigma\sigma'} = N g_{\sigma\sigma'}\Phi_\sigma \Phi_{\sigma'}$. An example of the excitation spectrum so obtained is shown in Fig.~\ref{fig:bose}(c). It's important to  note that once a specific condensate mode $\bPhi_l$ is chosen,  the BdG equation as well as the Bogoliubov amplitudes are also invariant under the $\calD_l$ rotation.  Within the Bogoliubov framework, the mass (spin) current operator becomes
$
\hat J^{(s)}_\mu \approx \sqrt{N}\sum_ {n\bk} \left [\calJ^{(s)}_{\mu,0n}(\bk)  \hat \alpha_{n\bk} + h.c.\right ]$, where $\hat \alpha_{n\bk}$ annihilates the quasi-particle with energy $\calE_{n\bk}$ and  $\calJ^{(s)}_{\mu,0n}(\bk) = \la \bPhi | j^{(s)}_\mu |\bu_{n\bk} \ra - \la \bv^*_{n\bk} | j^{(s)}_\mu |\bPhi\ra $ are (spin) current matrix elements.  From Eq.~(\ref{kubo}) we thus obtain the spin conductivity tensor for the Bose condensate as~\cite{SM}
\begin{equation}
    \sigma^{s}_{\mu\nu}=\frac{2}{A} \sum_{n\neq 0}\frac{{\rm Im} \left [\calJ^{s}_{\mu,0n}(0)\calJ^*_{\nu,0n}(0) \right ]}{(\calE_{n0}-\calE_{00})^2},
    \label{eq:SHall_w}
\end{equation}
 where we used the fact that $\calJ^{(s)}_{\mu,0n}(\bk)  = \delta_{\bk,0}\calJ^{(s)}_{\mu,0n}(0) $ due to quasi-momentum conservation.

Compared to the conventional intrinsic SHE exemplified by the degenerate Fermi gas previously discussed, the intrinsic SHE of the Bose condensate possesses a fundamental characteristic, i.e., the spin Hall conductivity is anisotropic due to the symmetry breaking caused by the in-plane magnetization. Recall that the condensate mode $\bPhi_l$ and the excitations break the $C_4$ symmetry but retain the $\calD_l$ symmetry. Under the $\calD_l$ rotation, the single particle current and spin current operator transform as $\calD_l j_x^{s}\calD_l^{-1}=(-1)^l j_{y}^{s}$ and $\calD_l j_y \calD_l^{-1}=(-1)^{l+1}j_{x}$ respectively. Applying the transformations to the matrix elements in Eq.~(\ref{eq:SHall_w}) we then find 
$
 \sigma_{xx}^{s}=-\sigma_{yy}^{s}$ and  $ \sigma_{xy}^{s}=-\sigma_{yx}^{s}$
for all four possible ground states $\bPhi_l$. Using these relations in Eq.~(\ref{scttr}) we arrive at 
\begin{align}
    \sigma_{y'x'}^{s}&=-\sin 2\theta \sigma_{xx}^{s}+\sigma_{yx}^{s}; \nn \\
    \sigma_{x'x'}^{s}&=\cos 2\theta \sigma_{xx}^{s}.
    \label{sigmabose}
\end{align}
In Fig.~\ref{fig:bose}(d), we have  explicitly verified Eq.~(\ref{sigmabose}) by calculating $\sigma^{s}_{y'x'}$ and $\sigma^{s}_{x'x'}$ using Eq.~(\ref{eq:SHall_w}) for all values of the rotation angle $\theta$ of $\hat \bx'$ relative to $\hat \bx$. We emphasize that the anisotropy in Eq.~(\ref{sigmabose}) is determined only by the symmetry properties of the ground state. This means that even though the value of the spin conductivity tensor depends on system parameters such as the cell-averaged density $\rho$ and the interaction strengths $g_{\sigma\sigma'}$, the relations in Eq.~(\ref{sigmabose}) remain the same as long as the condensate preserves the $\calD_l$ symmetry. Such a universality allows this unique anisotropy to be detected even in a harmonically trapped system where the cell-averaged density is no longer uniform.

\begin{figure}[tbh]
\begin{centering}
\includegraphics[width=8.6cm]{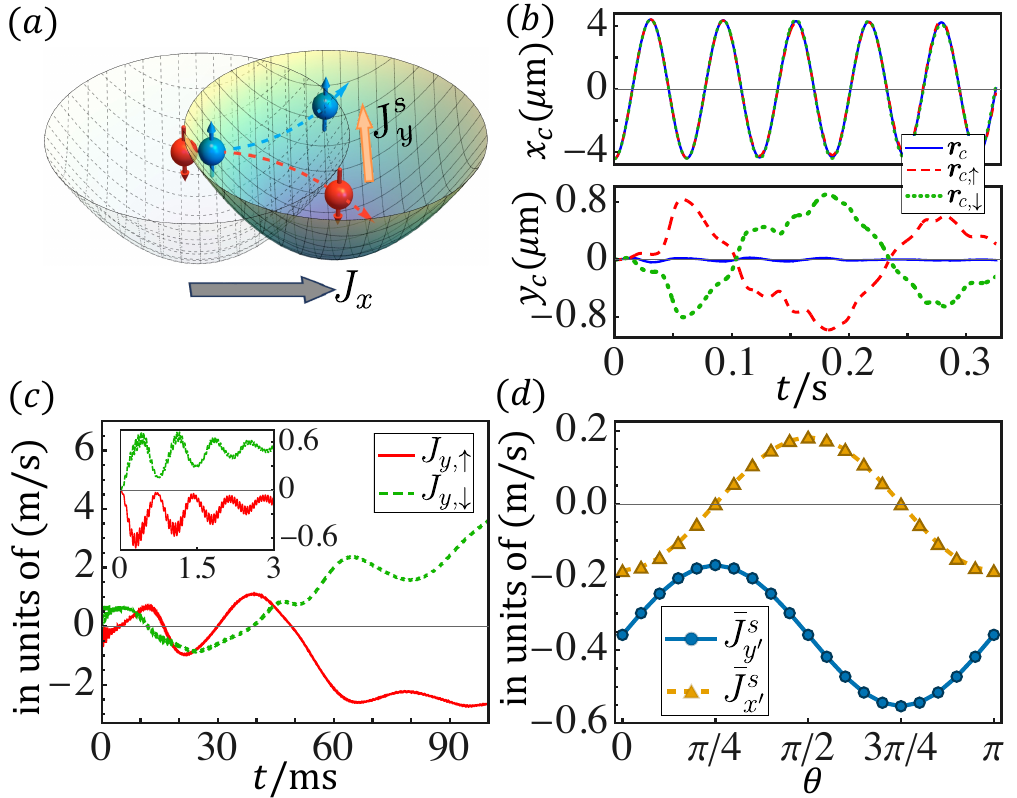}
\par\end{centering}
\caption{(a) Illustration of the detection proposal of the SHE. (b) Total COM motion and those of the different spin components after the trap is displaced by a distance of $d =4.43\mu$m along the $x$-direction. The center of the trap before displacement is set at $-d\hat \bx$ for convenience. (c) The transverse currents of the two spin components. (d) The spin currents $\bar J^s_{\mu}$ averaged over the first $3$ms as a function of the direction of the displacement. Here $k_L = 7.98\times 10^6$m$^{-1}$, $V_0 = 8E_r$, $M_0 = 1.5E_r$, $\omega_0 = 2\pi\times 30$Hz, $N \approx 3.2\times 10^5$,  $g_{\uparrow\uparrow} \approx 1.99\times 10^{-11}$Hz$\cdot$m$^2$, and $g_{\uparrow\downarrow} = 1.70\times 10^{-11}$Hz$\cdot$m$^2$.  }
\label{fig:GP}
\end{figure}

\emph{Detection proposal and numerical simulations.}---The experimental systems are trapped in an additional 2D harmonic potential $V_{\rm tr}(\br) = \frac{1}{2}m\omega_0^2 r^2$ which we have so far neglected. The presence of such a trap actually allows for a convenient probe of the SHE by means of dipole oscillations. Indeed, displacing the trap from $V_{\rm tr}(\br)$ to $\tilde V_{\rm tr}(\br)  = V_{\rm tr}(\br -\bd)  $ generates a force ${\boldsymbol F} = m\omega_0^2 \bd$ which, in turn, induces a mass current along the $\bd$ direction. According to the Ohm's law, an ensuing spin current along the transverse direction then demonstrates the SHE (see Fig.~\ref{fig:GP}(a)). 

To show that the unconventional SHE can be observed this way, we perform numerical simulations for the $^{87}$Rb condensate by varying the direction of the displacement $\bd$ from $\theta = 0$ to $2\pi$.  After  the trap displacement the condensate dynamics is described by the time-dependent GP equation~\cite{Kawaguchi2012}
\begin{align}
i{\pa_t \Phi_\sigma}= \sum_{\sigma'} [( h_{\sigma\sigma'}+   {\tilde V}_{\rm tr}\delta_{\sigma\sigma'})\Phi_{\sigma'}+Ng_{\sigma\sigma'}|\Phi_{\sigma'}|^2\Phi_\sigma ],
\end{align}
where $\bPhi(\br,t)$ is time-dependent condensate wave function. At $t=0 $ we have $\bPhi(\br, 0) = \bPhi_{\rm tr}$, where $\bPhi_{\rm tr}$ is the initial condensate mode for the trapped system obtained from Eq.~(\ref{GPequ}) after replacing the single-particle Hamiltonian $h $ by $h+V_{\rm tr} $~\cite{SM}. For concreteness, we specify $\bPhi_{\rm tr}$ as the state that is adiabatically connected to $\bPhi_1$ in Fig.~\ref{fig:bose} upon removing the trap. In Fig.~\ref{fig:GP}(a) and (b) we plot the dynamics of each component's center of mass (COM) $\br_{c,\sigma} = \int d\br \br |\Phi_\sigma(\br,t)|^2$ as well as that of the total COM $\br_c = \br_{c,\uparrow} +\br_{c,\downarrow} $ after the quench along the $x$-direction. Figure~\ref{fig:GP}(b) clearly shows that an out-of-phase oscillation between the two spin components along the $y$-direction accompanies the damped dipole oscillation along the $x$-direction, an unmistakable signature for the SHE. 

We can further calculate the spin currents $J^{s}_\mu =  (J_{\mu,\uparrow} -  J_{\mu,\downarrow})/2$, where $J_{\mu,\sigma} = (N/mi) \int d\br \Phi_\sigma^*(\br,t) \pa_\mu \Phi_\sigma(\br,t)$ (see for example $J_{y,\sigma}$ in Fig.~\ref{fig:GP}(c)). For a uniform system of area $A$, Ohm's law relates the spin currents to the spin Hall conductivity via $J^s_\mu/A = \sigma^s_{\mu\nu} F_\nu$, which allows one to extract $\sigma^s_{\mu\nu} $ from the time-averaged spin currents $\bar J^s_\mu$ measured in experiments. For a harmonically trapped system, however, the spin conductivity tensor $\sigma^s_{\mu\nu}$ depends on the local cell-averaged density $\rho(\br)$ and the Ohm's law becomes $J^s_\mu = \int d\br  \sigma^s_{\mu\nu} (\rho ) F_\nu$. Since the anisotropic behavior of $\sigma^s_{\mu\nu}$ in uniform systems does not depend on the value of the density, the angular dependence of the time-averaged spin currents for the trapped system must be exactly that of $\sigma^s_{\mu\nu}$ given in Eq.~(\ref{sigmabose}). As shown in Fig.~\ref{fig:GP}(d), this is precisely what we find when the simulation is repeated for trap displacement along all directions. This demonstrates the robustness of the unconventional SHE which is key to its experimental detection.

\emph{ Conclusions and outlook.---}We have shown that recently realized $\calPT$ symmetric spin-orbit coupled quantum gases are excellent platforms for studying the intrinsic SHE and related spin transport phenomena. In the case of the Fermi gas, we find that the relative importance of the geometric contribution to the spin Hall conductivity can be drastically different for insulating and for metallic states, providing insight into known mechanisms of the solid-state intrinsic SHE. For Bose condensates, we have predicted the existence of an unconventional coplanar SHE, which exhibits a unique anisotropic spin Hall conductivity due to the in-plane magnetization. Our numerical simulations of the unconventional SHE in a trapped system indicate that experimental verification of these findings is already within reach. Furthermore, our theory can also be applied to other $\calPT$ symmetric models that may potentially be realized using quantum gases. Finally, the inverse spin Hall effect, i.e., the generation of a charge current from a spin current, has never been demonstrated in quantum gases. We anticipate that this could be achieved by using spin-selective trapping potentials. A realistic proposal for its quantum simulation will be left for future work. 

We thank Yangqian Yan for helpful discussions. This work is supported by Natural Science Foundation of China (Grant No.~12474264, No.~12174455), Guangdong Provincial Quantum Science Strategic Initiative (Grant No.~GDZX2404007), National Key R$\&$D Program of China (Grant No. 2022YFA1404103), Natural Science Foundation of Guangdong
Province (Grant No. 2021B1515020026), and Guangdong
Basic and Applied Basic Research Foundation (Grant No.
2023B1515040023). S.Z. acknowledges support from HK GRF (Grant No. 17306024), CRF (Grants No. C6009-20G
and No. C7012-21G), and a RGC Fellowship Award No.
HKU RFS2223-7S03. 

\bibliography{SHE_REFS_v1}	

\end{document}


\title{Supplemental Material for ``Unconventional spin Hall effect in $\mathcal {PT}$ symmetric spin-orbit coupled quantum gases"}
\author{Hui Tang}
\altaffiliation{These authors contributed equally to this work.}
\affiliation{Shenzhen Institute for Quantum Science and Engineering, Southern University of Science and Technology, Shenzhen 518055, China}
\author{Guan-Hua Huang}
\altaffiliation{These authors contributed equally to this work.}
\affiliation{Hefei National Laboratory, Hefei 230088, China}
\author{Shizhong Zhang}
\affiliation{Department of Physics and HKU-UCAS Joint Institute for Theoretical and Computational Physics at Hong Kong, University of Hong Kong, Hong Kong, China}
\author{Zhongbo Yan}
\email{yanzhb5@mail.sysu.edu.cn}
\affiliation{Guangdong Provincial Key Laboratory of Magnetoelectric Physics and Devices, School of Physics, Sun Yat-sen University, Guangzhou 510275, China}
\author{Zhigang Wu}
\email{wuzhigang@quantumsc.cn}
\affiliation{Quantum Science Center of Guangdong-Hong Kong-Macao Greater Bay Area (Guangdong), Shenzhen 508045, China}
\date{\today }

\maketitle
This Supplemental Material includes the following two sections: (I) spin Hall conductivity of the Bose condensate and (II) numerical simulations of the unconventional SHE.

\section{Spin Hall conductivity of the Bose condensate}\label{sup:conductivity}
In this section, we provide more details on the derivation of  Eq.~(7) of the main text for the spin Hall conductivity of the Bose condensate in the in-plane magnetized phase, in the absence of the harmonic trap. 
We begin with the many-body Hamiltonian for $\mathcal P\mathcal T$ symmetric spin-orbit coupled Bose gas with anisotropic interactions
\begin{equation}
    \hat H = \int d^2\bm r\, \left[\sum_{\sigma}\hat\psi^\dagger_\sigma(\bm r) [h(\bm r)-\mu_{ B}]\hat \psi_\sigma(\bm r)+\frac{1}{2}\sum_{\sigma\sigma'}g_{\sigma\sigma'}\hat\psi_\sigma^\dagger(\bm r)\hat\psi_{\sigma'}^\dagger(\bm r)\hat\psi_{\sigma'}(\bm r)\hat\psi_\sigma(\bm r)\right], \label{eq:Htot}
\end{equation}
where $\mu_{B}$ is the chemical potential and the interaction strengths satisfy $g_{\uparrow\uparrow}=g_{\downarrow\downarrow}>g_{\uparrow\downarrow} $. The condensate wave function $\la \hat \bpsi \ra = \sqrt{N}\bPhi$, where $\hat\bpsi \equiv (\hat \psi_\uparrow, \hat \psi_\downarrow)^{T}$  and $\bPhi$ is normalized to unity, is determined by the Gross-Pitaevskii equation (5) of the main text. 
\begin{figure}[tb]
    \begin{centering}
    \includegraphics[width=16.4cm]{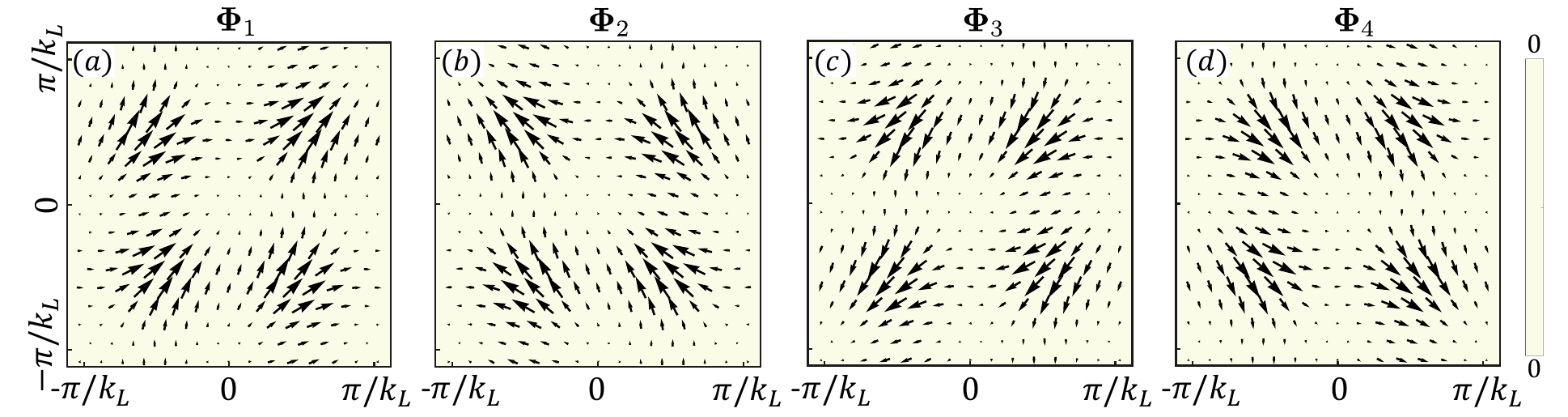}
    \par\end{centering}
    \caption{The magnetization density $\boldsymbol m_l (\br)$ of each $\bPhi_l$, shown in an area containing four lattice sites, where the uniform background color indicates a vanishing $z$-component of the magnetization. Here the parameters are the same as Fig.~2 of the main text, i.e., $V_0 = 8E_r$, $M_0  = 1.5E_r$, $\rho g_{\uparrow\uparrow} = 0.26 E_r $ and $\rho g_{\uparrow\downarrow} = 0.22 E_r$.}
    \label{fig:m}
\end{figure}
In the in-plane magnetized phase, the condensate wave function is a equal-weight supposition of the degenerate Bloch states at the $\Gamma$ point, i.e., 
\begin{align}
    \bPhi(\br) = \sum_{n} \left [c_n\bphi_{n0}(\br) + c_{\bar n} \bphi_{\bar n 0}(\br) \right ],
\end{align}
where $\bphi_{\bar n 0}(\br) = \mathcal{PT}\bphi_{n0}(\br)$ is the $\mathcal{PT}$ symmetric partner of $\bphi_{n0}(\br)$ and $|c_n| = |c_{\bar n}|$. For the system parameters considered by us, we find that $|c_0|^2 + |c_{\bar 0}|^2 > 0.99$ and so the atoms essentially condensate in the lowest two degenerate Bloch states. Furthermore, it turns out that the relative phase between these two states can take four possible values $\theta_l = l\pi/2 -\pi/4 $ ($l = 1,\cdots,4$), leading to a four-fold degeneracy in the condensate mode.  Namely, to a very good approximation, the four degenerate condensate modes are given by
\begin{align}
    \bPhi_l(\br) \approx \frac{1}{\sqrt{2}}\left [ \bphi_{00}(\br) + e^{i\theta_l} \bphi_{\bar 0 0}(\br) \right ].
\end{align}
Each of these four modes has a distinctive magnetization density defined as
\begin{align}
{\boldsymbol m}_l(\br) = \bPhi^\dag_l(\br) {\boldsymbol s} \bPhi_l(\br),
\end{align}
the integration of which yields the total magnetization
$
{\boldsymbol M}_l = \int d^2\br \,{\boldsymbol m}_l (\br)$.  The magnetization densities for these four modes are  shown in Fig.~\ref{fig:m}. 

To determine the elementary excitations, we substitute
\begin{align}
    \hat \psi_\sigma(\bm r)=\sqrt{ N}\Phi_\sigma(\bm r)+\delta\hat\psi_\sigma(\bm r)
    \label{psiexp}
    \end{align}
into the Hamiltonian \eqref{eq:Htot} to obtain Bogoliubov de Gennes (BdG) Hamiltonian
\begin{equation}
    \hat H_B = \int d^2\bm r\, \bm {\delta \hat \psi}^\dagger(\bm r) \mathcal{H}_B(\bm r)\bm{ \delta \hat\psi}(\bm r), \label{eq:Hb}
\end{equation}
where $\bm{\delta\hat \psi}=(\delta\hat\psi^T,\delta\hat\psi^\dagger)^T$ and
\begin{equation}
    \mathcal{H}_B(\bm r)=\pmat{h_0-\mu_B+\mathcal M & \mathcal N\\\mathcal N^* &h_0^*-\mu_B+ \mathcal M^*}.
\end{equation}
The Bogoliubov Hamiltonian in Eq.\,\eqref{eq:Hb} can be brought into a diagonalized form by the Bogoliubov transformation
\begin{equation}
    \delta\hat\psi_{\sigma}(\bm r)=\sum_{n\bm k}u_{n\bm k\sigma}(\bm r)\hat\alpha_{n\bm k}-v_{n\bm k\sigma}^*(\bm r)\hat\alpha_{n\bm k}^\dagger,
    \label{BdGtr}
\end{equation}
where $\bm k$ belongs to the first Brillouin zone and $n=0,1,2,...$ is the band index; $\hat\alpha_{n \bm k}^\dagger$ is the creation operator for the Bogoliubov quasiparticle, and $u_{n\bm k\sigma}$ and $v_{n\bm k\sigma}$ are the Bogoliubov amplitudes determined by Eq.~(6) of the main text. The Bogoliubov amplitudes satisfy the normalization condition 
\begin{align}
\int d^2\br [\bu^\dag_{n\bm k} (\br) \bu_{n\bm k} (\br)  - \bv^\dag_{n\bm k} (\br) \bv_{n\bm k} (\br) ]  = 1,
\end{align}
where $\bu_{n\bm k} = (u_{n\bm k\uparrow} ,u_{n\bm k\downarrow} )^T$ and $\bv_{n\bm k} = (v_{n\bm k\uparrow} ,v_{n\bm k\downarrow} )^T$.

To derive Eq.~(7) of the main text, we start from the definition
\begin{equation}
    \sigma^s_{\mu\nu}=-\lim_{\omega\rightarrow 0}\frac{1}{A\omega}{\rm Im}\chi_{\mu\nu}^s(\omega)
\end{equation}
where $A$ is system's area and $\chi_{\mu\nu}^s(\omega)$ is the Fourier transform of the retarded correlation function 
\begin{align}
\chi_{\mu\nu}^s(t-t') = -i\theta(t-t')\langle[\hat J_\mu^s(t),\hat J_\nu(t')]\rangle.
\end{align}
The current and spin current operator are respectively 
\begin{align}
    \hat{J}_\mu = & \frac{1}{2mi}\int d^2 \bm r\, \left[\hat\bpsi^\dagger (\bm r)\sigma_0\partial_\mu\hat \bpsi(\bm r) - h.c.\right]; \\
    \hat J^s_\nu =& \frac{1}{2mi}\int d^2 \bm r\, \left[\hat\bpsi^\dagger (\bm r)\frac{\sigma_z}{2}\partial_\nu\hat \bpsi(\bm r) - h.c.\right].
\end{align}
Substituting Eqs.~(\ref{psiexp}) and (\ref{BdGtr}) into the above equations, we find to the linear order in the quasi-particle operators
\begin{align}
    \hat{J}_\mu &\approx \sqrt{N}\sum_{n\bm k}\left[\mathcal{J}_{\mu,0n}(\bm k)\hat\alpha_{n\bm k}+h.c.  \right]; \\
    \hat J^s_\nu &\approx \sqrt{N}\sum_{n\bm k}\left[\mathcal{J}_{\mu,0n}^s(\bm k)\hat\alpha_{n\bm k}+h.c.  \right].
\end{align}
Here
\begin{align}
    \mathcal{J}^{(s)}_{\mu, 0n}(\bm k)& =\int d^2\br \left [  \bPhi^\dag(\br)  j_{\mu}^{(s)} \bu_{n\bk} (\br)  - \bv^T_{n\bk}(\br)  j_{\mu}^{(s)}  \bPhi(\br) \right ] \nonumber \\
    &\equiv \langle\bm\Phi|j_{\mu}^{(s)}|\bm u_{n\bm k}\rangle-\langle \bm v^*_{n\bm k}|j_{\mu}^{(s)}|\bm\Phi\rangle,
\end{align}
where the single-particle current operators are $ \bj = \bp/m$ and $ \bj^{s} = \frac{1}{2}\left \{ s_z, \bj \right \}$. Because of quasi-momentum conservation we in fact have 
\begin{align}
\mathcal{J}^{(s)}_{\mu, 0n}(\bm k) = \delta_{\bm k, 0}\mathcal{J}^{(s)}_{\mu, 0n}(0).
\end{align}
Using the spectrum representation and solutions of BdG equation, the spin Hall conductivity can then be expressed as
\begin{align}
    \sigma^s_{\mu\nu} &= -\lim_{\omega \rightarrow 0}\frac{1}{A \omega} {\rm Im}\sum_{n\neq 0}\left[\frac{\mathcal J^s_{\mu ,0n}(0)\mathcal J_{\nu,n0}(0)}{\omega-\mathcal E_{n0}+\mathcal E_{00}+i0^+}-\frac{\mathcal J_{\nu,0n}(0)\mathcal J_{\mu,n0}^s(0)}{\omega+\mathcal E_{n0}-\mathcal E_{00}+i0^+}  \right] \nonumber \\
    & = \frac{2}{A} \sum_{n\neq 0}\frac{{\rm Im} \left [\calJ^{s}_{\mu,0n}(0)\calJ^*_{\nu,0n}(0) \right ]}{(\calE_{n0}-\calE_{00})^2}.
    \label{eq:SHall_w}
\end{align}


\section{Numerical simulations of the unconventional SHE}
In this section we provide more details on the numerical simulations of the unconventional SHE in a harmonically trapped Bose condensate. In particular, we describe the preparation of the initial state and the condensate dynamics associated with the SHE. Our numerical simulations are performed for a quasi-2D two-component $^{87}$Rb gas with $N \approx 3.2\times 10^5$ atoms confined in a harmonic trap with frequency $\omega_0 = 2\pi\times 30$Hz~\cite{PhysRevLett.121.150401}. Other relevant system parameters are $k_L = 7.98\times 10^6$m$^{-1}$, $V_0 = 8E_r$, $M_0 = 1.5E_r$,   $g_{\uparrow\uparrow} \approx 1.99\times 10^{-11}$Hz$\cdot$m$^2$ and $g_{\uparrow\downarrow} = 1.70\times 10^{-11}$Hz$\cdot$m$^2$. For real space discretization, we use a $3000\times 3000$ grid with an area of about $37\mu$m $\times$ $37\mu$m. For the time evolution of the Gross-Pitaevskii (GP) equation, the four-order splitting operator method is implemented in order to achieve high precision~\cite{PhysRevA.60.1956, bao2012mathematical}. 

\begin{figure}[tb]
	\centering
	\includegraphics[width=16.0cm,height=6.18cm]{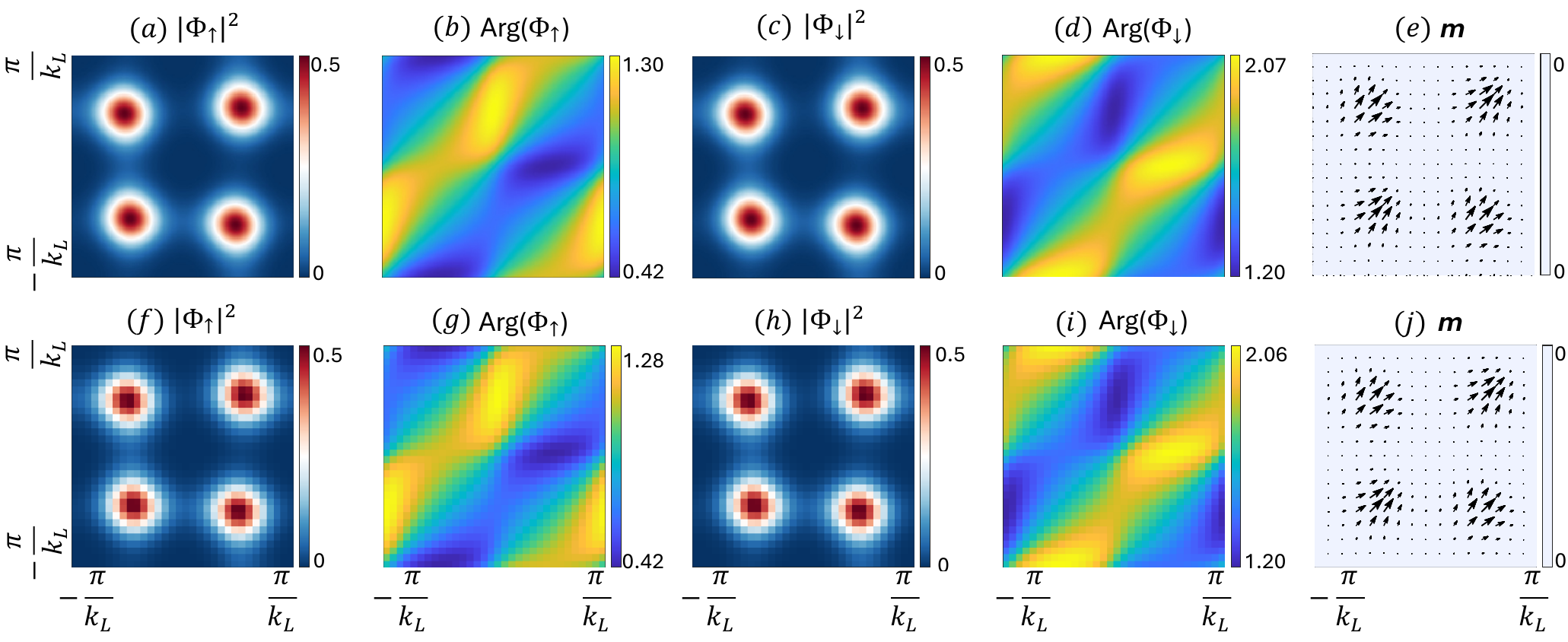}
		\caption{The initial condensate wave function $\bPhi_{\rm tr}$ of the trapped system at the trap center in an area containing four lattice sites (lower panel) compared to that of the uniform system $\bPhi_1$ of the same area (upper panel).  }
		\label{fig:state}
	\end{figure}
    \begin{figure}[h]
	\centering
	\includegraphics[width=15cm]{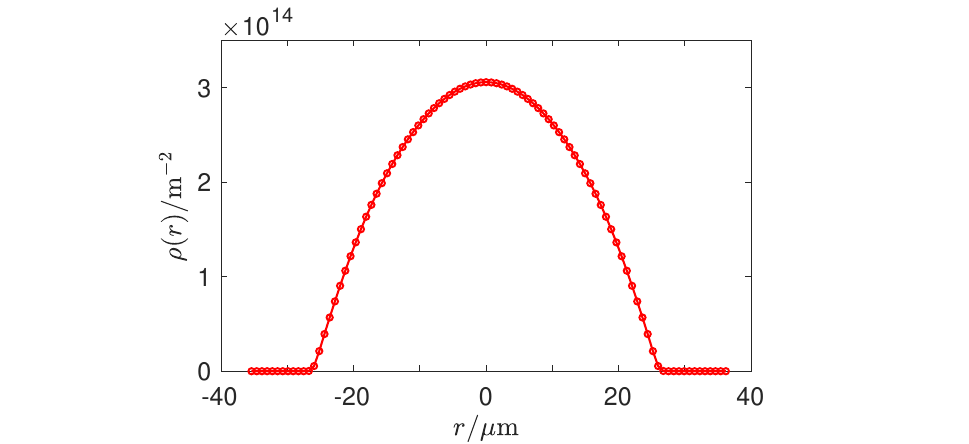}
		\caption{The cell-averaged density $\rho(r)$ of the trapped condensate, where $r$ denotes the distance of a unit cell from the origin. }
		\label{fig:cadensity}
	\end{figure}
We prepare the system in the ground state $\bPhi_{\rm tr}$, obtained from solving the time-independent GP equation
\begin{align}
\sum_{\sigma'}\left \{ [h_{\sigma\sigma'}(\br)+V_{\rm tr}(\br)\delta_{\sigma\sigma'}]\Phi_{\sigma'}+Ng_{\sigma\sigma'}|\Phi_{\sigma'}|^2\Phi_{\sigma}\right \}=\mu_{B}\Phi_{\sigma}
\label{TIGP}
\end{align}
using the method of imaginary time propagation. Since there is a four-fold degeneracy in the condensate mode, for concreteness, we choose the ground state that is adiabatically connected to $\bPhi_1$ of the uniform system when we switch off the trap. In fact, we have made sure that system parameters at the center of the trap are identical to those of the uniform system in the main text, i.e., at the center of the trap we have $\rho g_{\uparrow\uparrow} = 0.26 E_r$ and $\rho g_{\uparrow\downarrow} = 0.22 E_r$. Thus the condensate wave function of the trapped system at the center can be benchmarked against that of the uniform system. This comparison is shown in Fig.~\ref{fig:state}, where we find excellent agreement between these two. This agreement is one of the criteria we use to check the convergence of the ground state. For a global view of the condensate, we have plotted in Fig.~\ref{fig:cadensity} the cell-averaged density $\rho(r)$ corresponding to this initial state of the trapped condensate. 

\begin{figure}[bht]
	\centering
	\includegraphics[width=15.75cm,height=15.5cm]{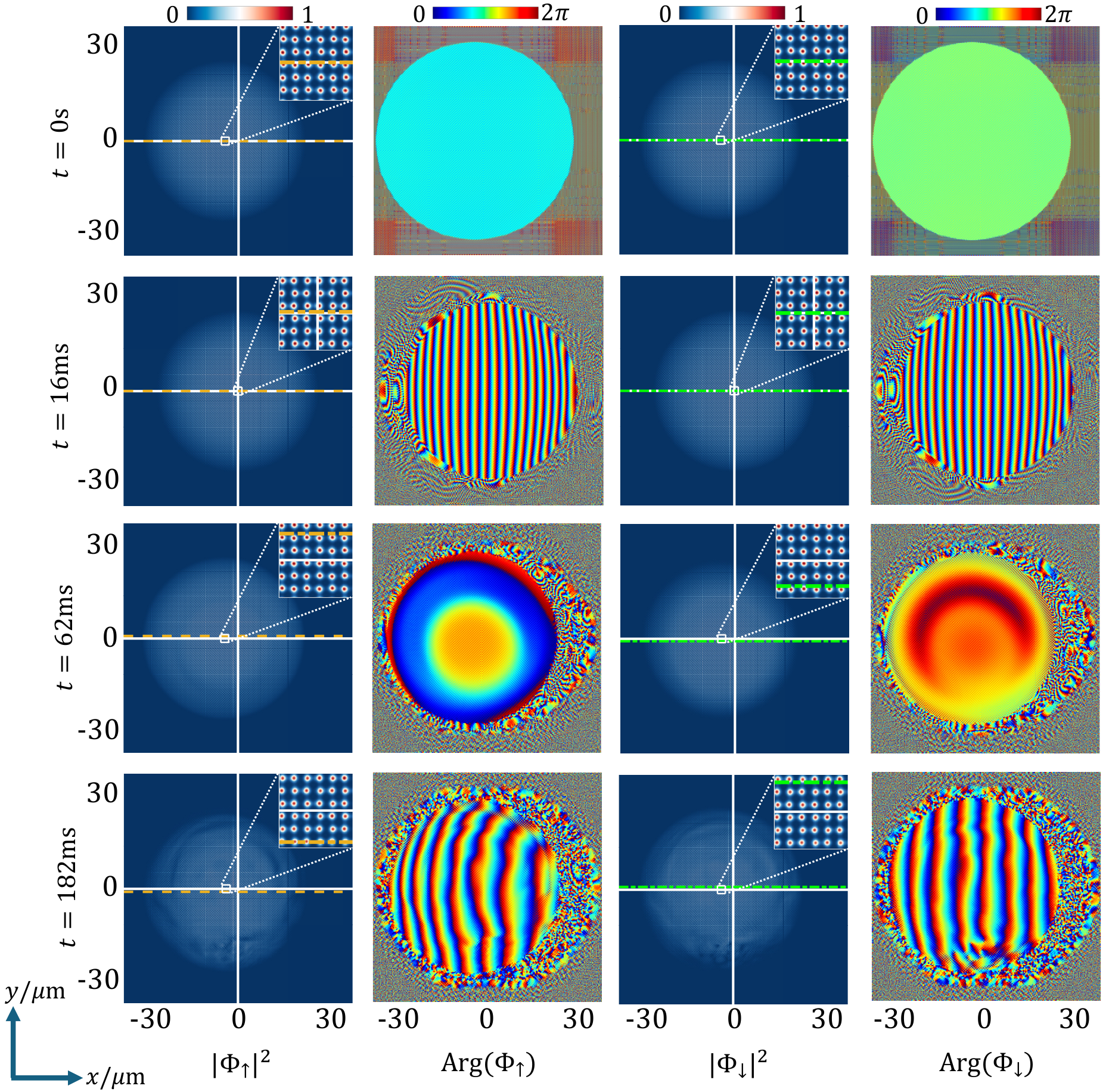}
	\caption{The densities and the phases of the condensate wave functions for the two spin components at various times after the displacement of the trap. The horizontal dashed lines indicate the $y$-coordinates of the center of mass of the spin components.}
	\label{fig:realGP}
\end{figure}
To probe the SHE, we displace the trap by a distance of $d= 4.43\mu$m along the $x$-direction and  numerically evolve the time-dependent GP equation 
\begin{align}
i{\partial_t \Phi_\sigma}(\br,t)= \sum_{\sigma'} [( h_{\sigma\sigma'}+   {\tilde V}_{\rm tr}(\br)\delta_{\sigma\sigma'})\Phi_{\sigma'}(\br,t)+Ng_{\sigma\sigma'}|\Phi_{\sigma'}|^2\Phi_\sigma(\br,t) ]
\end{align}
from the initial state $\bPhi(\br,0) = \bPhi_{\rm tr}$ determined above. For convenience, we have placed the center of the initial trap $V_{\rm tr}(\br)$ at $x = -d$ so that the center of the displaced trap $\tilde V_{\rm tr}(\br)$ is now at the origin, i.e., $V_{\rm tr}(\br) = \frac{1}{2}m\omega_0^2 (\br + d\hat\bx)^2$ and $\tilde V_{\rm tr}(\br) = \frac{1}{2}m\omega_0^2 r^2$.  Examples of the condensate wave function $\bPhi(\br, t)$ at various times are shown in Fig.~\ref{fig:realGP}. These wave functions are then used to calculate the center of mass dynamics of the spin components as well as the currents shown in Fig.~3a and Fig.~3b of the main text. Finally, the simulation is repeated for trap displacements along all directions to determine the angular dependence of the spin Hall currents shown in Fig.~3d. 

\bibliography{BibSISM.bib}